\documentclass[sigconf,nonacm,balance=false]{acmart}
\setcopyright{none}
\acmConference{}{}{}
\acmBooktitle{}
\acmDOI{}
\acmISBN{}
\DeclareUnicodeCharacter{2011}{\nobreakdash-}

\usepackage{tikz}
\usepackage{amsmath}
\usepackage{amsfonts}
\usepackage{enumitem}
\usepackage{algorithm}
\usepackage{algpseudocode}
\usepackage{tabularx}
\usepackage{booktabs}
\usepackage{graphicx}
\usepackage{textcomp}
\usepackage{multirow} 
\usepackage{makecell}
\usepackage{url}
\usepackage{xspace}
\usepackage{pifont}   % for \ding
\usepackage[english]{babel}

\newcommand{\cmark}{\ding{51}} % check mark
\newcommand{\xmark}{\ding{55}} % cross mark
\definecolor{comment}{rgb}{0.54,0.1,0.066}
\newcommand{\sys}{\textsf{NetLexicon}\xspace}

\definecolor{algcomment}{gray}{0.35}
\definecolor{designborder}{gray}{0.72}
\definecolor{designhead}{gray}{0.86}
\definecolor{designbody}{gray}{0.97}
\newif\ifshowchanges
\showchangesfalse       % 关闭颜色，恢复正常黑色

\newcommand{\rev}[1]{%
  \ifshowchanges
    \textcolor{blue}{#1}%
  \else
    #1%
  \fi
}
\begin{document}

\title{NetLexicon: Learning Discrete Behavioral Representations for Encrypted Web Traffic Analysis}

% Author order and affiliations supplied by the authors.
\author{Xiangyu Gao}
\affiliation{\institution{Institute for Network Sciences and Cyberspace, Tsinghua University}\city{Beijing}\country{China}}
\affiliation{\institution{State Key Laboratory of Internet Architecture}\country{China}}
\author{Tong Li}
\affiliation{\institution{Renmin University of China}\city{Beijing}\country{China}}
\affiliation{\institution{State Key Laboratory of Internet Architecture}\country{China}}
\author{Ziqiang Wang}
\affiliation{\institution{Department of Computer Science and Technology, Tsinghua University}\city{Beijing}\country{China}}
\author{Yinchao Zhang}
\affiliation{\institution{Department of Computer Science and Technology, Tsinghua University}\city{Beijing}\country{China}}
\affiliation{\institution{State Key Laboratory of Internet Architecture}\country{China}}
\author{Rongbang Wu}
\affiliation{\institution{Renmin University of China}\city{Beijing}\country{China}}
\author{Zhenxing Zhang}
\affiliation{\institution{Huawei}\country{China}}
\author{Jing Hu}
\affiliation{\institution{Huawei}\country{China}}
\author{Hanlin Huang}
\affiliation{\institution{Department of Computer Science and Technology, Tsinghua University}\city{Beijing}\country{China}}
\author{Xinle Du}
\affiliation{\institution{Department of Computer Science and Technology, Tsinghua University}\city{Beijing}\country{China}}
\author{Su Yao}
\affiliation{\institution{Beijing National Research Center for Information Science and Technology (BNRist), Tsinghua University}\city{Beijing}\country{China}}
\affiliation{\institution{State Key Laboratory of Internet Architecture}\country{China}}
\author{Qi Li}
\affiliation{\institution{Institute for Network Sciences and Cyberspace, Tsinghua University}\city{Beijing}\country{China}}
\author{Ke Xu}
\affiliation{\institution{Department of Computer Science and Technology, Tsinghua University}\city{Beijing}\country{China}}

\renewcommand{\shortauthors}{Gao et al.}
\authorsaddresses{}

% Compact preprint author block; individual author metadata is retained above.
\makeatletter
\renewcommand{\@mkauthors}{%
  \begingroup
  \global\setbox\mktitle@bx=\vbox{%
  \hsize=\textwidth
  \unvbox\mktitle@bx\par\medskip
  \centering
  {\normalfont\fontsize{11}{15}\selectfont
    Xiangyu Gao\textsuperscript{1,6}\quad
    Tong Li\textsuperscript{2,6}\quad
    Ziqiang Wang\textsuperscript{3}\quad
    Yinchao Zhang\textsuperscript{3,6}\quad
    Rongbang Wu\textsuperscript{2}\\[4pt]
    Zhenxing Zhang\textsuperscript{4}\quad
    Jing Hu\textsuperscript{4}\quad
    Hanlin Huang\textsuperscript{3}\quad
    Xinle Du\textsuperscript{3}\quad
    Su Yao\textsuperscript{5,6}\quad
    Qi Li\textsuperscript{1}\quad
    Ke Xu\textsuperscript{3}\par}
  \vspace{9pt}
  {\normalfont\fontsize{9}{12}\selectfont
    \textsuperscript{1}Institute for Network Sciences and Cyberspace, Tsinghua University\qquad
    \textsuperscript{2}Renmin University of China\\[3pt]
    \textsuperscript{3}Department of Computer Science and Technology, Tsinghua University\qquad
    \textsuperscript{4}Huawei\\[3pt]
    \textsuperscript{5}BNRist, Tsinghua University\qquad
    \textsuperscript{6}State Key Laboratory of Internet Architecture\par}
  \bigskip
  }%
  \endgroup}
\makeatother

\begin{abstract}
Encrypted Web traffic analysis requires effective representations of observable communication behavior. Existing pretraining methods often adapt NLP/CV objectives and sequence architectures, motivating learning objectives that capture traffic-specific interaction patterns. We present \sys, a discrete pretraining framework that learns reusable behavioral states from unlabeled traffic. It converts contextual traffic windows into discrete states through vector quantization, constructing a compact traffic lexicon. We design two complementary pretraining objectives. State Transition Prediction (STP) forecasts subsequent sequence structure and packet features from observed history, while Statistical Feature Alignment (SFA) grounds learned states in window-level traffic statistics. Together, they guide the lexicon to capture recurring communication behaviors and their evolution.

We evaluate \sys on four benchmarks covering Web application identification, service type identification, and malware detection. \sys improves Macro-F1 by up to 25.5 percentage points over the strongest baseline on each benchmark and reduces fine-tuning time per epoch by up to 23.6$\times$ relative to the evaluated baselines. Further analysis shows that the learned discrete states capture recognizable patterns in packet size, timing, and data transfer. These results demonstrate that incorporating observable behavioral structure into pretraining supports effective, efficient, and interpretable representations for encrypted traffic analysis.
\end{abstract}

\maketitle

% Neutral preprint running heads.
\fancyhead[LO,LE]{\small NetLexicon}
\fancyhead[RO,RE]{\small Gao et al.}

\section{Introduction}
\label{sec:intro}

Web activities such as page retrieval and content delivery produce observable traffic patterns even under encryption. Encrypted traffic analysis uses observable packet lengths, directions, and timing to identify websites, applications, and potential threats, supporting Web service management, network measurement, and security~\cite{nguyen2009survey,finsterbusch2013survey}. To support these tasks, supervised models learn to associate observed traffic patterns with application or activity labels, making labeled examples central to their training and adaptation. High-quality labels are expensive to obtain and maintain, while evolving services, applications, and network environments cause distribution shifts that require repeated model updates~\cite{wickramasinghe2025sok}. This motivates learning reusable representations from abundant unlabeled traffic to reduce downstream labeling requirements and adaptation costs.

\begin{figure}[t]
  \centering
  \includegraphics[width=\linewidth]{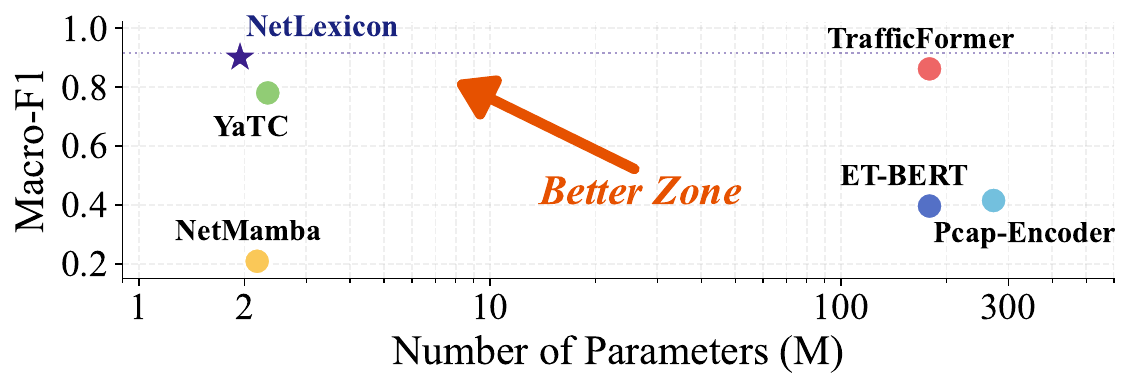}
  \caption{Model size and classification performance on CipherSpectrum~\cite{wickramasinghe2025sok}. \sys achieves the highest Macro-F1 with under 2M parameters.}
  \label{fig:moti}
  \vspace{-1.7em}
\end{figure}

Self-supervised pretraining offers a promising approach~\cite{devlin2019bert,he2022masked,et-bert,zhao2023yet}. Inspired by natural language processing (NLP) and computer vision (CV), recent methods learn traffic representations through tokenization, serialization, and masked prediction and reconstruction objectives, while exploring larger or more complex backbones~\cite{et-bert,zhao2023yet,wang2024netmamba,zhou2025trafficformer,zhao2025sweet,chen2025miett}. Figure~\ref{fig:moti} compares performance and model size on CipherSpectrum~\cite{wickramasinghe2025sok}, where \sys achieves the highest Macro-F1 with under 2M parameters. This motivates a closer look at objective design: how can traffic-intrinsic behavioral characteristics support efficient learning of meaningful communication representations?

In this paper, \textbf{traffic semantics refers to the patterns of data transfer and interaction produced by application activities and protocol exchanges, such as concentrated responses following requests, sustained bulk transfer, and intermittent exchanges of small amounts of data.} Packet lengths characterize exchange volume, inter-arrival times capture transmission rhythms, and direction changes and burst structures describe endpoint interactions~\cite{nguyen2009survey,moore2005internet,iscxvpn}. For example, a short upstream burst followed by large downstream packets in Web traffic may reflect content delivery following a request, while small bidirectional exchanges separated by longer intervals may reflect lightweight background interactions. Their combination and evolution within flow context provide a direct basis for learning traffic semantics.

Existing methods have advanced input organization and sequence modeling. As summarized in Table~\ref{tab:work-comparison}, prior approaches incorporate flow-level context and traffic-specific objectives to varying degrees, while explicitly grounding learned representations in observable behavior remains underexplored. Learning reusable traffic semantics calls for objectives that connect two complementary properties: temporal dependencies make the current interaction predictive of subsequent transmissions, while recurring packet-length, direction, and timing patterns provide observable characteristics of shared behaviors across flows. These observations motivate our central question: \textbf{how can self-supervised traffic pretraining abstract encrypted interaction patterns into reusable behavioral states and learn their evolution over time?}

We introduce \sys\footnote{\rev{\url{https://anonymous.4open.science/r/Open-NetLexicon/}}}, a discrete pretraining framework that encodes packet and burst sequences within flow-level context using header fields and metadata such as packet sizes and timing. It maps window representations to a finite set of codewords through vector quantization~\cite{van2017neural}, constructing a \emph{traffic lexicon}. Similar window-level behaviors share states, whose sequence describes how communication evolves.

\begin{table}[t]
  \centering
  \footnotesize
  \setlength{\tabcolsep}{4pt}
  \renewcommand{\arraystretch}{1.1}
  \caption{Comparison of traffic pretraining designs.}
  \label{tab:work-comparison}

  \begin{tabularx}{\linewidth}{@{}l
      >{\centering\arraybackslash}X
      >{\centering\arraybackslash}X
      >{\centering\arraybackslash}X
      >{\centering\arraybackslash}X@{}}
    \toprule
    \textbf{Method} &
    \textbf{Header Only} &
    \textbf{Flow Context$^{1}$} &
    \textbf{Traffic Objective$^{2}$} &
    \textbf{Behavior Grounding$^{3}$} \\
    \midrule
    ET-BERT~\cite{et-bert}
      & \xmark & \cmark & \cmark & \xmark \\
    YaTC~\cite{zhao2023yet}
      & \xmark & \cmark & \xmark & \xmark \\
    NetMamba~\cite{wang2024netmamba}
      & \xmark & \cmark & \xmark & \xmark \\
    TrafficFormer~\cite{zhou2025trafficformer}
      & \xmark & \cmark & \cmark & \xmark \\
    Pcap-Encoder~\cite{zhao2025sweet}
      & \cmark & \xmark & \cmark & \xmark \\
    \sys~(Ours)
      & \textbf{\cmark} & \textbf{\cmark}
      & \textbf{\cmark} & \textbf{\cmark} \\
    \bottomrule
  \end{tabularx}

  \parbox{\linewidth}{\raggedright\footnotesize
  $^{1}$ The model encodes multiple packets to capture inter-packet dependencies.\\
  $^{2}$ The pretraining task is tailored to traffic characteristics.\\
  $^{3}$ Representations are aligned with observable traffic behavior.}
  \vspace{-1.5em}
\end{table}

To learn states that capture how traffic evolves and what behavior each window exhibits, we design two self-supervised pretraining objectives: \textbf{State Transition Prediction (STP)} and \textbf{Statistical Feature Alignment (SFA)}. \textbf{STP} predicts subsequent sequence structure and packet features from observed context, encouraging the learned states to preserve information about future communication behavior. \textbf{SFA} aligns each quantized state with packet-length, direction, and inter-arrival-time statistics from the same window, grounding the state in observable behavioral characteristics. Together, our objectives shape the traffic lexicon through temporal prediction and statistical alignment, enabling it to capture recurring behaviors and their evolution.

We pretrain \sys on backbone and QUIC/HTTP3 application traffic~\cite{mawi,visquic} and evaluate it on four tasks. Website fingerprinting on CSTNET~\cite{et-bert} and encrypted application fingerprinting on CipherSpectrum~\cite{wickramasinghe2025sok} assess its ability to distinguish Web-related communication behaviors. Service type identification on ISCXVPN~\cite{iscxvpn} and malware detection on USTC-TFC~\cite{ustc} assess its broader applicability. We further examine interpretability through associations between representations and behavioral statistics, codebook utilization, and codeword cases, alongside training and inference overhead.

We make the following contributions:
\begin{itemize}[leftmargin=*,itemsep=0.2em,topsep=0.2em]
\item \textbf{Reframing traffic pretraining.}
We connect packet lengths, inter-arrival times, and directions to interaction patterns, organizing encrypted traffic representations through shared behavioral states and their evolution.

\item \textbf{Learning a discrete traffic lexicon.}
We introduce \sys, which combines vector quantization with our STP and SFA objectives to learn predictive and behaviorally grounded discrete states from unlabeled traffic.

\item \textbf{Effectiveness, efficiency, and interpretability.}
Across four Web-related and broader traffic classification tasks, \sys improves Macro-F1 by up to 25.5 percentage points, reduces fine-tuning time per epoch by up to 23.6$\times$, and learns interpretable states aligned with meaningful communication behaviors.
\end{itemize}
\section{Background and Related Work}
\label{sec:background}

We briefly introduce the necessary background and related work to situate our approach.

\subsection{Traffic Representation Learning}

Traffic classification identifies applications, services, and malicious activities from observed traffic flows, supporting network measurement and security. Dynamic ports and payload encryption have driven the use of statistical features, including packet size distributions, inter-arrival times, and flow statistics~\cite{nguyen2009survey,finsterbusch2013survey,dpi,middlebox}. Conventional approaches combine these features with classifiers such as SVMs and random forests~\cite{moore2005internet,nguyen2009survey}. Deep learning methods further learn representations from packet and flow data using CNNs, RNNs, and GNNs~\cite{lotfollahi2020deep,liu2019fs,wang2020app,shen2021accurate,aceto2019mimetic,pean}, with applications to website fingerprinting, encrypted stream identification, and mobile traffic analysis~\cite{sirinam2018deep,rimmer2017automated,schuster2017beauty,razaghpanah2015haystack,TLSAndroid}. These developments establish representation learning as an important component of traffic analysis.

Self-supervised pretraining learns transferable representations by constructing supervision from unlabeled data and adapting the pretrained encoder to labeled downstream tasks~\cite{devlin2019bert,he2022masked,chen2020simple,he2020momentum}. Two common approaches are sequence modeling, which predicts masked or future observations, and contrastive learning, which aligns related samples and separates unrelated ones~\cite{InfoNCE,chen2020simple}. Traffic pretraining methods build on these ideas through masked byte modeling in ET-BERT, masked autoencoder reconstruction in YaTC, and sequence modeling with architectures such as NetMamba and TrafficFormer~\cite{et-bert,zhao2023yet,wang2024netmamba,zhou2025trafficformer}. Other studies further explore relationships between packets and flows, multimodal information, and representations built around bursts~\cite{chen2025miett,mm4flow2025,flowmae}. Together, these methods provide a foundation for learning from large traffic corpora and adapting to downstream classification tasks.

Recent studies emphasize the importance of what these representations learn. Pcap-Encoder revisits pretraining objectives based on packet headers, while CipherSpectrum and related critiques show how dataset artifacts, spurious correlations, and evaluation practices affect encrypted traffic classification~\cite{zhao2025sweet,wickramasinghe2025sok}. Broader studies in security machine learning also highlight shortcut learning and data leakage as concerns for reliable evaluation~\cite{arp2022and,jacobs2022ai}. These findings motivate pretraining objectives grounded in observable communication behavior. For \sys, this means learning representations that preserve packet size, direction, and timing patterns together with their temporal dependencies, providing a behavioral foundation for adaptation across traffic analysis tasks.

\vspace{-1em}
\subsection{Learning Discrete Traffic States}

Network traffic exhibits recurring patterns shaped by protocol behavior and communication dynamics, including packet structure, timing, bursts, and relationships across packets~\cite{mm4flow2025}. These patterns motivate a representation of traffic as transitions among latent behavioral states, such as request bursts, sustained responses, and intermittent exchanges. In \sys, a traffic state denotes a learned behavioral category for traffic windows that captures recurring communication patterns across flows. Grouping similar windows into shared states provides a compact way to organize these patterns, while state sequences describe how interactions evolve. This perspective connects the choice of representation to the behavioral semantics available in encrypted traffic~\cite{wickramasinghe2025sok}.

\begin{figure*}[t]
  \centering
  \includegraphics[width=\textwidth]{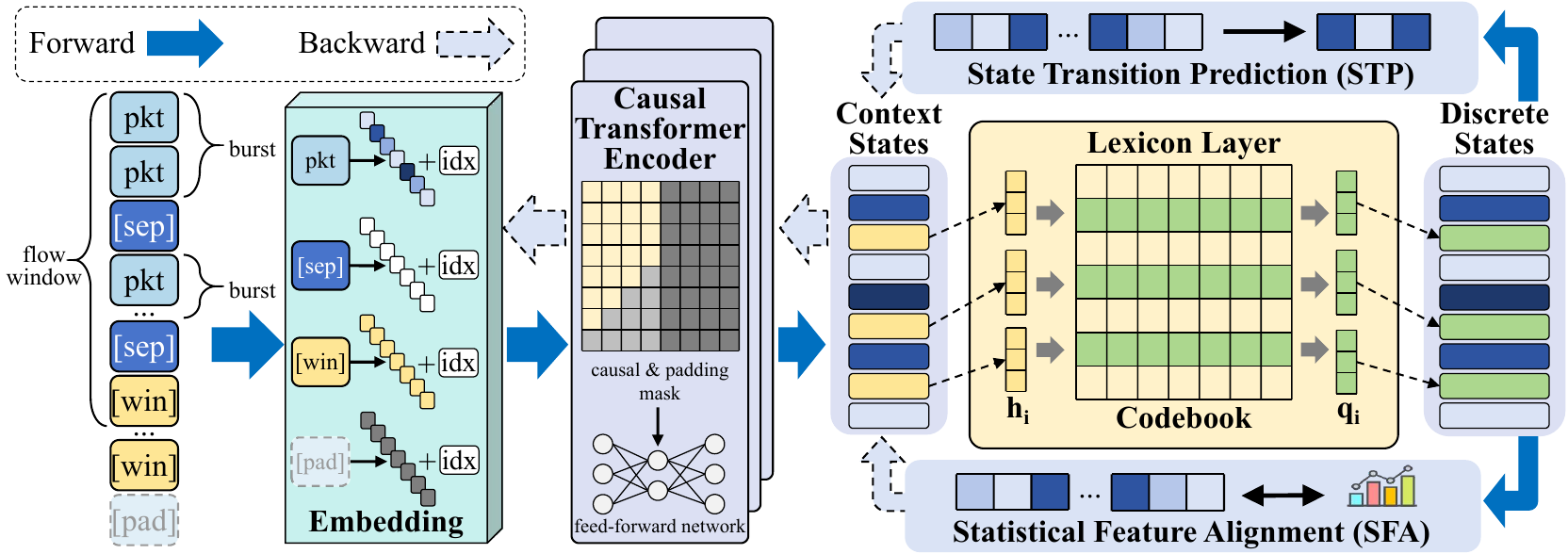}
  \caption{Overview of \sys.}
  \label{fig:overview}
  \vspace{-1em}
\end{figure*}

Learning discrete latent variables requires handling nondifferentiable assignments during optimization~\cite{jang2016categorical,maddison2016concrete}. We build on Vector Quantized Variational Autoencoders (VQ-VAE), which map each encoder output to its nearest entry in a learned codebook and use the corresponding index as a discrete latent variable~\cite{van2017neural}. This produces a finite vocabulary of reusable representations~\cite{van2017neural,esser2021taming}. In our setting, codewords represent behavioral states of traffic windows. Related traffic models also exploit burst and flow structure. Flow-MAE uses preprocessing based on bursts and learns from fixed byte patches, while TrafficFormer represents directional bursts as tokens for masked modeling~\cite{flowmae,zhou2025trafficformer}. Building on these traffic organization cues, \sys quantizes contextual window representations into a discrete traffic lexicon. We design State Transition Prediction and Statistical Feature Alignment to shape this lexicon through future behavior prediction and alignment with observable window statistics.

\vspace{-1em}
\section{NetLexicon}
\label{sec:design}

\noindent\textbf{Goal.}
\sys aims to learn reusable traffic semantics from unlabeled flows by abstracting the interaction patterns of Web activities into discrete behavioral states. These states capture how data is exchanged and how communication evolves, supporting downstream traffic analysis.

\noindent\textbf{Challenges.}
Packet lengths, directions, and timing jointly reflect communication behavior, whose interpretation depends on the surrounding flow context. Learning a shared state vocabulary requires preserving these temporal dependencies while associating each state with observable characteristics of recurring interactions across flows.

\noindent\textbf{Solutions.}
\sys encodes packet and burst sequences within flow context and quantizes window representations into a discrete traffic lexicon. We design State Transition Prediction (STP) to capture how interactions evolve and Statistical Feature Alignment (SFA) to characterize the behavior each window exhibits. Together, they guide the lexicon toward predictive states grounded in measurable traffic patterns. Figure~\ref{fig:overview} presents the overall framework.

\vspace{-1em}
\subsection{Framework}

The input to \sys is a bidirectional flow defined by source IP, destination IP, source port, destination port, and transport protocol, with packets ordered chronologically. Following TrafficFormer~\cite{zhou2025trafficformer}, a burst contains consecutive packets in the same direction, and a window contains \(W\) consecutive bursts, with \(W=5\) by default. We insert a \texttt{[SEP]} token after each burst and a \texttt{[WIN]} token after every \(W\) bursts. These markers expose interaction boundaries within the packet sequence.

Each packet token contains seven normalized features and seven binary missing indicators, yielding a 14-dimensional input. The features are packet size, direction, inter-arrival time, TCP flags, IPID delta, relative TCP sequence number, and relative TCP acknowledgement number. Table~\ref{tab:feature_design} summarizes the selection, which emphasizes observable communication behavior and relative transport progress. \texttt{[PAD]} tokens are zero filled and masked out from attention and all pretraining losses.

Packet features, token types, and positions are embedded into a shared hidden space and processed by a causal transformer. The output at position \(t\) uses observations up to \(t\). At each valid \texttt{[WIN]} position, the encoder therefore summarizes the traffic prefix through the current window. We quantize this representation into a discrete state. STP supervises the encoded sequence through predictions at valid token positions, while SFA aligns quantized window states with statistics from their corresponding windows.

\begin{table}[!t]
\centering
\caption{Feature choices in \sys.}
\label{tab:feature_design}
\small
\renewcommand{\arraystretch}{1.05}
\begin{tabular*}{\linewidth}{@{\extracolsep{\fill}}lccc@{}}
\toprule
Feature group & Used in model & Easy shortcut & Behavior$^{1}$ \\
\midrule
Packet size & \checkmark & \xmark & \checkmark \\
Direction & \checkmark & \xmark & \checkmark \\
Interarrival time & \checkmark & \xmark & \checkmark \\
TCP flags & \checkmark & \xmark & \checkmark \\
\(\Delta\)IPID & \checkmark & \xmark & \checkmark \\
Raw IPID & \xmark & \checkmark & \xmark \\
Rel.\ SEQ & \checkmark & \xmark & \checkmark \\
Abs.\ SEQ & \xmark & \checkmark & \xmark \\
Rel.\ ACK & \checkmark & \xmark & \checkmark \\
Abs.\ ACK & \xmark & \checkmark & \xmark \\
IP address & \xmark & \checkmark & \xmark \\
Port number & \xmark & \checkmark & \xmark \\
TTL & \xmark & \checkmark & \xmark \\
\bottomrule
\end{tabular*}

\parbox{\linewidth}{\raggedright\footnotesize
$^{1}$ Indicates whether the feature primarily reflects communication behavior and state evolution, without encoding endpoint identity or arbitrary absolute values.}
\vspace{-1.5em}
\end{table}

\subsection{Discrete Traffic Lexicon}

The traffic lexicon groups contextual windows into a finite set of reusable behavioral states. Quantization encourages recurring patterns to share codewords, while causal encoding provides the packet, burst, and historical context needed to distinguish their roles in an interaction.

Let \(\mathbf{h}_i \in \mathbb{R}^{d}\) denote the representation at the boundary of window \(i\), and let \(\mathcal{E}=\{\mathbf{e}_k\}_{k=1}^{K}\) be a learned codebook with \(\mathbf{e}_k \in \mathbb{R}^{d}\). The state index is assigned by nearest-neighbor lookup
\begin{equation}
z_i = \arg\min_{k \in \{1,\ldots,K\}}
\|\mathbf{h}_i-\mathbf{e}_k\|_2^2,
\label{eq:vq_assign}
\end{equation}
and its quantized representation is
\begin{equation}
\mathbf{q}_i = \mathbf{e}_{z_i}.
\label{eq:vq_quantized}
\end{equation}
Across a flow, these assignments produce a state sequence \(\{z_i\}\) and its corresponding representations \(\{\mathbf{q}_i\}\).

We update the codebook through exponential moving averages (EMA) of assignment counts and assigned representations, as shown in Algorithm~\ref{alg:codebook_learning}. A straight-through estimator passes gradients from quantized representations to encoder outputs. The commitment loss keeps the outputs close to their assigned codewords
\begin{equation}
\mathcal{L}_{\mathrm{LEX}}
=
\frac{1}{M}
\sum_{i=1}^{M}
\left\|
\mathbf{h}_i-\operatorname{sg}[\mathbf{q}_i]
\right\|_2^2,
\label{eq:vq_commit}
\end{equation}
where \(\operatorname{sg}[\cdot]\) is the stop-gradient operator and \(M\) is the number of valid windows in the batch.

The default codebook contains \(K=128\) entries, with EMA decay \(0.99\) and stability constant \(10^{-5}\). Every 500 steps, entries with EMA assignment counts below \(0.01\) are checked for reinitialization to maintain codebook utilization. Section~\ref{sec:evaluation} examines sensitivity to codebook and window sizes.

\begin{algorithm}[t]
\caption{EMA based codebook update in \sys}
\label{alg:codebook_learning}
\begin{algorithmic}[1]
\Require Window representations \(\{\mathbf{h}_i\}_{i=1}^{M}\),
codebook \(\mathcal{E}=\{\mathbf{e}_k\}_{k=1}^{K}\),
EMA states \(\{N_k\}_{k=1}^{K}\) and
\(\{\mathbf{m}_k\}_{k=1}^{K}\),
decay rate \(\gamma\), stability constant \(\epsilon\)
\Ensure Assignments \(\{z_i\}_{i=1}^{M}\) and updated codebook \(\mathcal{E}\)

\For{each \(\mathbf{h}_i\)}
    \State \(z_i \gets \arg\min_k
    \|\mathbf{h}_i-\mathbf{e}_k\|_2^2\)
\EndFor

\For{each codeword index \(k\)}
    \State \(n_k \gets
    \sum_{i=1}^{M}\mathbf{1}[z_i=k]\)
    \State \(\mathbf{s}_k \gets
    \sum_{i=1}^{M}\mathbf{1}[z_i=k]\,\mathbf{h}_i\)
    \State \(N_k \gets
    \gamma N_k+(1-\gamma)n_k\)
    \State \(\mathbf{m}_k \gets
    \gamma\mathbf{m}_k+(1-\gamma)\mathbf{s}_k\)
    \State \(\mathbf{e}_k \gets
    \mathbf{m}_k/\max(N_k,\epsilon)\)
\EndFor

\State Periodically reinitialize underutilized codewords
\State \Return \(\{z_i\}\) and \(\mathcal{E}\)
\end{algorithmic}
\end{algorithm}

\subsection{State Transition Prediction}

\begin{figure}[t]
  \centering
  \includegraphics[width=\linewidth]{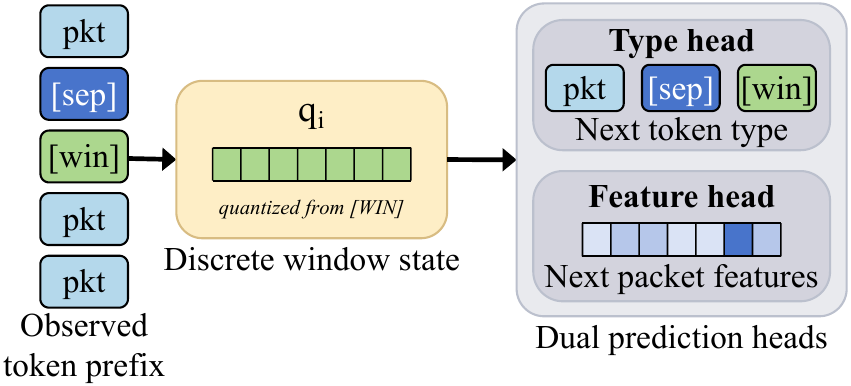}
  \caption{An illustration of the State Transition Prediction task.}
  \label{fig:stp}
\end{figure}

We design STP to make representations predictive of how communication evolves. Request bursts, responses, and waiting periods provide context for subsequent packets and interaction boundaries. A useful window state should therefore summarize the observed interaction and preserve information about what follows. STP uses this temporal dependency as supervision through two prediction heads, as illustrated in Figure~\ref{fig:stp}.

Following causal self-attention~\cite{vaswani2017attention}, position \(t\) attends to positions up to \(t\) and predicts a target at \(t+1\). We construct the targets by shifting the input sequence by one position. The heads operate at valid positions, using quantized states at \texttt{[WIN]} positions and contextual encoder outputs elsewhere. Predictions at window boundaries thus train the discrete states to capture future behavior, while predictions at other positions provide dense supervision for the shared encoder.

The first head predicts whether the next token is a packet, a burst separator, or a window separator. This task captures whether the sequence continues within a burst or reaches an interaction boundary. Let \(y_t^{\mathrm{type}} \in \{\texttt{pkt},\texttt{[SEP]},\texttt{[WIN]}\}\) denote the target and \(\hat{\mathbf{p}}_t^{\mathrm{type}}\) the predicted distribution. The loss is
\begin{equation}
\mathcal{L}_{\mathrm{type}}
=
-\frac{1}{|\mathcal{T}|}
\sum_{t \in \mathcal{T}}
\log \hat{\mathbf{p}}_t^{\mathrm{type}}
\big(y_t^{\mathrm{type}}\big),
\label{eq:stp_type}
\end{equation}
where \(\mathcal{T}\) contains valid positions whose next token is also valid.

When the next token is a packet, the second head predicts its normalized feature vector, including packet size, direction, timing, and transport information. Let \(\mathcal{P}\subseteq\mathcal{T}\) contain these positions, with target \(\mathbf{y}_t^{\mathrm{feat}}\in\mathbb{R}^{d_f}\) and prediction \(\hat{\mathbf{y}}_t^{\mathrm{feat}}\), where \(d_f=7\). The feature loss is
\begin{equation}
\mathcal{L}_{\mathrm{feat}}
=
\frac{1}{|\mathcal{P}|}
\sum_{t \in \mathcal{P}}
\left\|
\hat{\mathbf{y}}_t^{\mathrm{feat}}
-
\mathbf{y}_t^{\mathrm{feat}}
\right\|_2^2.
\label{eq:stp_feat}
\end{equation}
Together, the type head learns the structural progression of the sequence, and the feature head learns the characteristics of the packets that follow. Their combined objective is
\begin{equation}
\mathcal{L}_{\mathrm{STP}}
=
\lambda_{\mathrm{type}}\mathcal{L}_{\mathrm{type}}
+
\lambda_{\mathrm{feat}}\mathcal{L}_{\mathrm{feat}},
\label{eq:stp_total}
\end{equation}
where \(\lambda_{\mathrm{type}}\) and \(\lambda_{\mathrm{feat}}\) balance the two losses.

\subsection{Statistical Feature Alignment}

\begin{figure}[t]
  \centering
  \includegraphics[width=\linewidth]{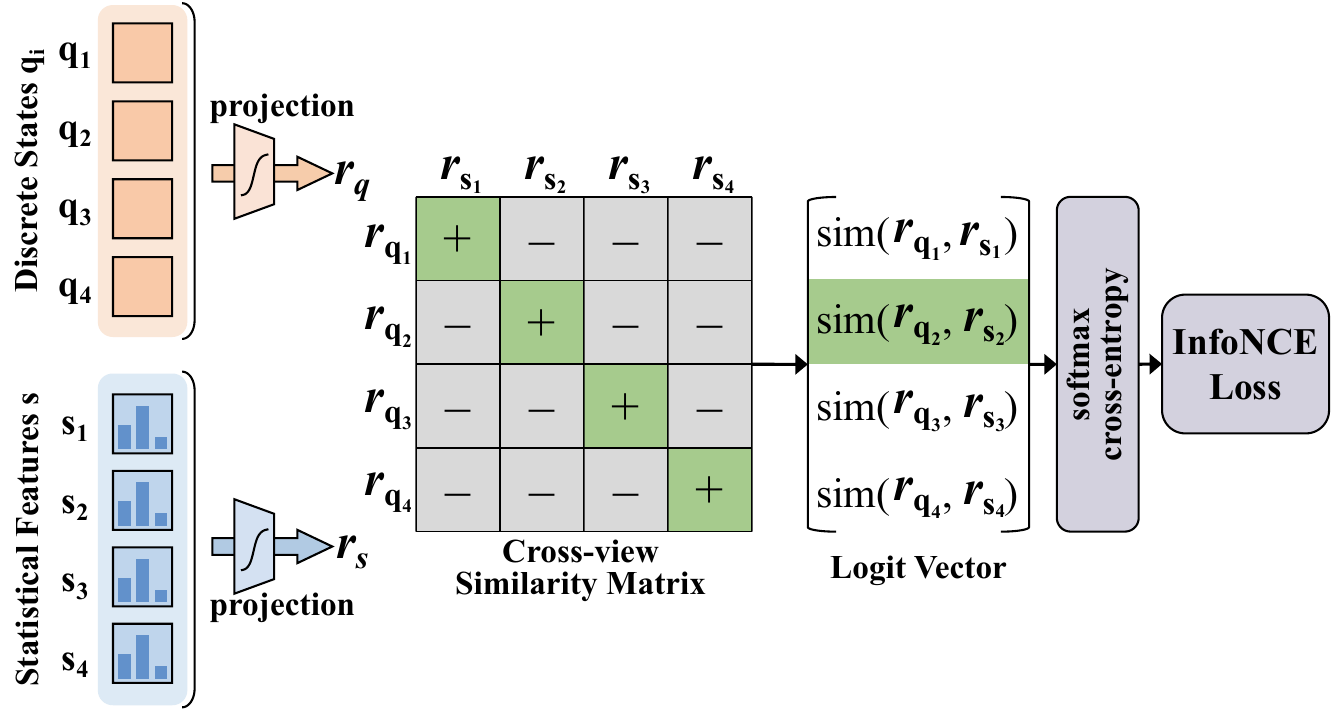}
  \caption{Loss computation of Statistical Feature Alignment.}
  \label{fig:sfa}
\end{figure}

We design SFA to associate each discrete state with measurable characteristics of its traffic window. STP encourages a state to predict subsequent behavior, while SFA preserves information about the behavior exhibited within the current window. It provides this supervision by aligning two views of the same communication segment, as shown in Figure~\ref{fig:sfa}.

The first view is the quantized state \(\mathbf{q}_i\), and the second is a statistical vector \(\mathbf{f}_i\in\mathbb{R}^{d_s}\) computed from the same window. We use \(d_s=37\) features summarizing packet and byte volume, directional balance, packet length and timing distributions, TCP control flags, and active or idle behavior. These statistics provide supervision directly from observed traffic. The state \(\mathbf{q}_i\) incorporates preceding flow context, while \(\mathbf{f}_i\) summarizes the current window. Aligning them anchors the contextual state to the behavior observed in that segment.

Two learnable heads project the state and statistical views into a shared contrastive space
\begin{equation}
\mathbf{r}_i^{q}=g_q(\mathbf{q}_i), \qquad
\mathbf{r}_i^{s}=g_s(\mathbf{f}_i).
\label{eq:sfa_proj}
\end{equation}
Both heads output 64-dimensional vectors, which are normalized before comparison. Within a batch of \(M\) valid windows, \((\mathbf{r}_i^{q},\mathbf{r}_i^{s})\) forms a positive pair. Pairing the same state view with statistics from another window, \((\mathbf{r}_i^{q},\mathbf{r}_j^{s})\) for \(j\neq i\), forms a negative pair. Using cosine similarity \(\mathrm{sim}(\cdot,\cdot)\), SFA minimizes the InfoNCE loss~\cite{InfoNCE}
\begin{equation}
\mathcal{L}_{\mathrm{SFA}}
=
-\frac{1}{M}
\sum_{i=1}^{M}
\log
\frac{
\exp\left(\mathrm{sim}(\mathbf{r}_i^{q},\mathbf{r}_i^{s})/\tau\right)
}{
\sum_{j=1}^{M}
\exp\left(\mathrm{sim}(\mathbf{r}_i^{q},\mathbf{r}_j^{s})/\tau\right)
},
\label{eq:sfa_loss}
\end{equation}
with temperature \(\tau=0.07\). The loss encourages each state view to match its corresponding statistical view more closely than those of other windows. This guides quantized states to preserve the statistical characteristics of their assigned windows, giving the traffic lexicon an observable behavioral basis.

\subsection{Pretraining Objective and Adaptation}

The full pretraining objective combines temporal prediction, codebook commitment, and statistical alignment
\begin{equation}
\mathcal{L}
=
\mathcal{L}_{\mathrm{STP}}
+
\lambda_{\mathrm{LEX}}\mathcal{L}_{\mathrm{LEX}}
+
\lambda_{\mathrm{SFA}}\mathcal{L}_{\mathrm{SFA}},
\label{eq:pretrain_total}
\end{equation}
where \(\lambda_{\mathrm{LEX}}=0.25\) and \(\lambda_{\mathrm{SFA}}=0.1\) by default. All supervision is constructed from unlabeled traffic through subsequent token types, packet features, and window statistics. The objectives jointly train the shared encoder and associated prediction and projection heads. Gradients pass through quantization to the encoder, while the codebook follows the EMA update in Algorithm~\ref{alg:codebook_learning}.

For downstream tasks, \sys initializes the embedding layer and encoder with pretrained parameters and pools the contextual \texttt{[WIN]} representations into a flow representation. A task-specific classification head maps this representation to downstream labels. Fine-tuning uses cross-entropy loss on labeled flows, adapting the pretrained behavioral representations to the target task.
\section{Evaluation}
\label{sec:evaluation}

We evaluate \sys across four downstream tasks to examine classification performance, the contribution of its pretraining components, and the behavioral information captured by the discrete traffic lexicon. We further compare model size, fine-tuning time, inference latency, and memory usage to assess its cost.

\vspace{-0.7em}
\subsection{Experimental Setup}
\label{sec:exp_setup}

\noindent\textbf{Pretraining Data.}
Our pretraining corpus combines MAWI~\cite{mawi} and VisQUIC~\cite{visquic}. MAWI provides diverse backbone traffic from the WIDE trans-Pacific link, while VisQUIC contributes encrypted QUIC/HTTP3 application traffic from modern Web environments. Their combination covers both backbone communication and application interactions across TCP and UDP. We use SplitCap~\cite{splitcap} to extract bidirectional flows following the definition in Section~\ref{sec:design}. The resulting corpus contains 1,057,331 flows and approximately 86.34 million packets, with 820,986 flows from MAWI and 236,345 from VisQUIC.

\noindent\textbf{Downstream Data and Splits.}
Table~\ref{tab:finetune} summarizes the four downstream datasets after preprocessing. CSTNET supports web application fingerprinting across 120 classes using TLS~1.3 traffic. CipherSpectrum also targets web application fingerprinting and balances each application class across three TLS~1.3 cipher suites, reducing correlations between application labels and cipher configurations. These datasets assess fine-grained discrimination in web application related settings. ISCXVPN covers service categories under VPN conditions, and USTC-TFC contains normal application and malware traffic, extending evaluation to broader traffic analysis tasks.

We apply the flow and burst preprocessing described in Section~\ref{sec:design} and construct sliding windows of five consecutive bursts. Flows containing fewer than one complete window are discarded. Each dataset is randomly partitioned within each class at the flow level into training, validation, and test sets with an 8:1:1 ratio and seed 42. All windows from the same flow remain in the same split. The resulting flow partitions are shared across methods.

\begin{table}[t]
\centering
\caption{Downstream datasets after preprocessing.}
\label{tab:finetune}
\small
\setlength{\tabcolsep}{3pt}
\begin{tabularx}{\linewidth}{@{}l X r r@{}}
\toprule
Dataset & Task & Flows & Classes \\
\midrule
CSTNET~\cite{et-bert}
  & Web APP fingerprinting & 45.16K & 120 \\
CipherSpectrum~\cite{wickramasinghe2025sok}
  & Web APP fingerprinting & 164.2K & 42 \\
ISCXVPN~\cite{iscxvpn}
  & Service type identification & 2.184K & 7 \\
USTC-TFC~\cite{ustc}
  & Malware detection & 114.3K & 17 \\
\bottomrule
\end{tabularx}
\vspace{-1.5em}
\end{table}

\begin{table*}[t]
\centering
\caption{Performance comparison on Service Type Identification and Malware Detection.}
\label{tab:main_results_easy}
\setlength{\tabcolsep}{10pt}
\resizebox{\textwidth}{!}{%
\begin{tabular}{l cccc cccc}
\toprule
\multirow{2}{*}{Method}
  & \multicolumn{4}{c}{ISCXVPN~\cite{iscxvpn}}
  & \multicolumn{4}{c}{USTC-TFC~\cite{ustc}} \\
\cmidrule(lr){2-5}\cmidrule(lr){6-9}
  & Accuracy & Precision & Recall & F1-Score
  & Accuracy & Precision & Recall & F1-Score \\
\midrule
ET-BERT~\cite{et-bert} 	 	 	
  & 0.4178 & 0.3665 & 0.3018 & 0.2850 
  & 0.9681 & 0.9073 & 0.8956 & 0.9011 \\
YaTC~\cite{zhao2023yet}
  & 0.6000 & 0.5985 & 0.5274 & 0.5373
  		
  & 0.9799 & 0.9531 & \textbf{0.9627} & 0.9523 \\
NetMamba~\cite{wang2024netmamba}
  & 0.4044 & 0.3405 & 0.3965 & 0.3321	 
  & 0.5333 & 0.6914 & 0.4114 & 0.4215 \\
TrafficFormer~\cite{zhou2025trafficformer}
  & 0.6044 & 0.5566 & 0.5452 & 0.5485
  		
  & 0.9753 & \textbf{0.9754} & 0.9169 & 0.9215 \\
Pcap-Encoder~\cite{zhao2025sweet}
  & 0.5411 & 0.5117 & 0.5413 & 0.5124	
  & 0.5465 & 0.4830 & 0.5820 & 0.5041 \\
\midrule
\sys~(Ours)
  & \textbf{0.8267} & \textbf{0.8100} & \textbf{0.8003} & \textbf{0.8034}
  & \textbf{0.9889} & 0.9658 & 0.9474 & \textbf{0.9547} \\
\bottomrule
\end{tabular}%
}
\end{table*}

\begin{table*}[t]
\centering
\caption{Performance comparison on Web Application Fingerprinting.}
\label{tab:main_results_hard}
\setlength{\tabcolsep}{10pt}
\resizebox{\textwidth}{!}{%
\begin{tabular}{l cccc cccc}
\toprule
\multirow{2}{*}{Method}
  & \multicolumn{4}{c}{CSTNET~\cite{et-bert}}
  & \multicolumn{4}{c}{CipherSpectrum~\cite{wickramasinghe2025sok}} \\
\cmidrule(lr){2-5}\cmidrule(lr){6-9}
  & Accuracy & Precision & Recall & F1-Score
  & Accuracy & Precision & Recall & F1-Score \\
\midrule
ET-BERT~\cite{et-bert}
  & 0.6039 & 0.5944 & 0.5396 & 0.5335
  & 0.4440 & 0.3695 & 0.4514 & 0.3961 \\
YaTC~\cite{zhao2023yet}
  & 0.7356 & 0.7283 & 0.7069 & 0.7030
  & 0.8022 & 0.7908 & 0.7894 & 0.7799 \\
NetMamba~\cite{wang2024netmamba}
  & 0.7146 & 0.7233 & 0.6849 & 0.6836
  & 0.2665 & 0.2260 & 0.2602 & 0.2088 \\
TrafficFormer~\cite{zhou2025trafficformer}
  & 0.7616 & 0.7473 & 0.7287 & 0.7322
  & 0.8650 & 0.8606 & 0.8633 & 0.8614 \\
Pcap-Encoder~\cite{zhao2025sweet} 
  & 0.4180 & 0.4010 & 0.3950 & 0.3980 
  & 0.4360 & 0.4210 & 0.4090 & 0.4149 \\
\midrule
\sys~(Ours)
  & \textbf{0.8838} & \textbf{0.8757} & \textbf{0.8614} & \textbf{0.8637}
  & \textbf{0.9104} & \textbf{0.9049} & \textbf{0.8999} & \textbf{0.9002} \\
\bottomrule
\end{tabular}%
}
\end{table*}

\noindent\textbf{Training Settings.}
We implement \sys in PyTorch 2.5.1 and conduct experiments on NVIDIA A100 GPUs. The encoder contains four causal transformer layers with hidden size 192, feed-forward size 768, 16 attention heads, and a maximum sequence length of 256. The codebook contains 128 states, and the pretraining objectives follow Section~\ref{sec:design}. We pretrain for 20 epochs on one GPU with batch size 128, using AdamW with a learning rate of \(2\times10^{-4}\), cosine decay, and 1,000 warmup steps. The pretraining checkpoint is selected by validation loss.

Fine-tuning follows two stages. We first freeze the backbone and train the classification head at ten times the base learning rate, then unfreeze the model for end-to-end optimization. The base learning rate is \(5\times10^{-5}\), and the batch size is 128. We use AdamW with weight decay and gradient clipping, and optimize cross-entropy loss. The fine-tuning checkpoint is selected by validation accuracy, with the test split evaluated after checkpoint selection.

\noindent\textbf{Baselines.}
We compare \sys with ET-BERT~\cite{et-bert}, YaTC~\cite{zhao2023yet}, NetMamba~\cite{wang2024netmamba}, TrafficFormer~\cite{zhou2025trafficformer}, and Pcap-Encoder~\cite{zhao2025sweet}. These methods cover byte sequence modeling, flow image reconstruction, alternative sequence backbones, and header-based pretraining. We retain their public pretrained checkpoints and use consistent downstream partitions, fine-tuning settings, checkpoint-selection criteria, and evaluation metrics wherever applicable.

Each baseline retains its input representation. ET-BERT and TrafficFormer use five packets with 64 bytes per packet. YaTC and NetMamba use \(40\times40\) flow images. Pcap-Encoder operates on packet-level samples in our reproduction. Thus, the comparison evaluates adaptation from each method's existing pretrained model under the shared downstream protocol.

\noindent\textbf{Metrics.}
We report accuracy, macro-precision, macro-recall, and Macro-F1 for classification, with Macro-F1 as the primary metric because it gives equal weight to each class. Behavioral analysis examines codebook utilization, associations between representation distances and traffic statistics, and representative codeword profiles. Efficiency measurements cover parameter count, fine-tuning time per epoch, inference latency, and GPU memory usage.

\vspace{-0.7em}
\subsection{Traffic Classification}

Tables~\ref{tab:main_results_easy} and~\ref{tab:main_results_hard} report classification results across all four tasks. \sys achieves the highest accuracy and Macro-F1 on each dataset. On CSTNET, it reaches 86.37\% Macro-F1 across 120 web application classes, exceeding the strongest baseline by 13.15 percentage points. On CipherSpectrum, it achieves 90.02\% Macro-F1, improving over TrafficFormer by 3.88 points. The latter result shows that the advantage remains when application classes are balanced across cipher configurations. Together, these two benchmarks demonstrate the usefulness of the learned representation for fine-grained encrypted traffic classification in web application settings.

The results also extend to other traffic analysis tasks. On ISCXVPN, \sys improves Macro-F1 from 54.85\% to 80.34\%, a gain of 25.49 percentage points under VPN conditions. On USTC-TFC, the strongest baseline already reaches 95.23\%, and \sys achieves a smaller improvement to 95.47\%. The different gain sizes reflect the varying difficulty and baseline performance of the tasks.

\begin{figure}[t]
  \centering
  \includegraphics[width=\linewidth]{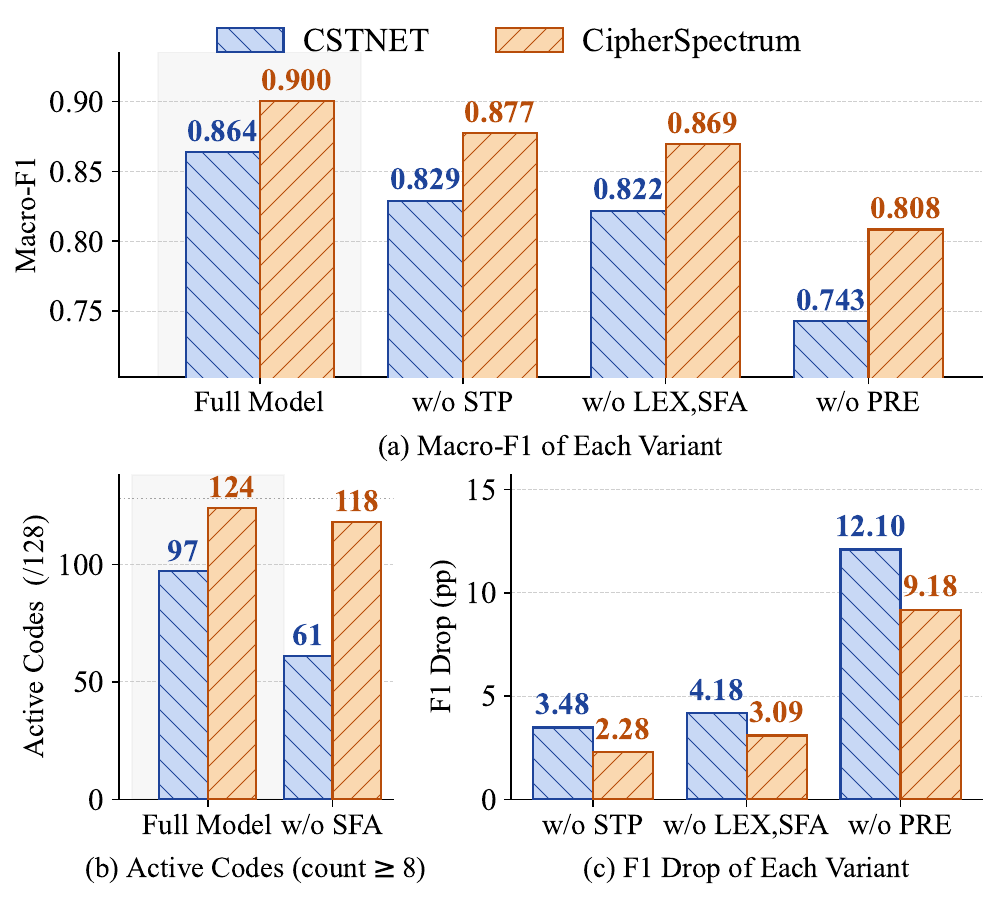}
  \caption{Component ablation on CSTNET and CipherSpectrum, showing Macro-F1, codebook utilization, and performance drops.}
  \label{fig:ablation}
  \vspace{-1.5em}
\end{figure}

\noindent\textbf{Component Ablation.}
Figure~\ref{fig:ablation} examines the contribution of the main components on CSTNET and CipherSpectrum. The full model achieves the highest Macro-F1 among the evaluated variants. Removing STP reduces performance, supporting the value of temporal supervision through future packet and sequence prediction. Jointly removing LEX and SFA also reduces performance, demonstrating the contribution of the discrete lexicon together with statistical alignment. Removing SFA decreases the number of active codes from 97 to 61 on CSTNET and from 124 to 118 on CipherSpectrum, indicating that alignment with window statistics encourages broader codebook utilization. Removing pretraining causes the largest Macro-F1 drops, reaching 12.10 and 9.18 percentage points on the two datasets, respectively, highlighting the importance of the initialization learned from unlabeled traffic.

\noindent\textbf{Parameter Sensitivity.}
We also examine codebook and window sizes through independently trained configurations on CipherSpectrum. For \(K\in\{32,64,128,256,512\}\), test accuracy ranges from 90.95\% to 91.16\%, indicating limited sensitivity to codebook capacity in this range. Both \(K=64\) and \(K=128\) achieve an active-code ratio of 98.4\%. Window size introduces a tradeoff between statistical context and flow coverage. Although \(W=3\) achieves higher classification performance, 47.96\% of its windows contain only three or four packets, providing limited context for window statistics. Increasing \(W\) to 7 and 9 reduces usable short-flow coverage to 85.6\% and 68.9\%, respectively. The default \(W=5\) balances these considerations. These sensitivity results were obtained after fixing the default configuration and were not used for its selection.
\vspace{-0.5em}
\subsection{Traffic Semantics Analysis}

We examine whether the learned representations preserve observable communication behavior through a quantitative comparison with baselines and an analysis of representative codewords. On the CipherSpectrum test set, 124 of the 128 codes receive at least eight assignments, showing broad use of the available state vocabulary.

\noindent\textbf{Association with Observable Behavior.}
We focus on packet length and inter-arrival time, which characterize the scale and pacing of data exchange. Both are established signals in statistical traffic classification, encrypted stream identification, and VPN traffic analysis~\cite{nguyen2009survey,moore2005internet,schuster2017beauty,iscxvpn}. Full-flow mean packet size and \(\log(1+\mathrm{mean\ IAT})\) serve as common probes for comparing the representations learned by different methods.

\begin{figure}[t]
  \centering
  \includegraphics[width=\linewidth]{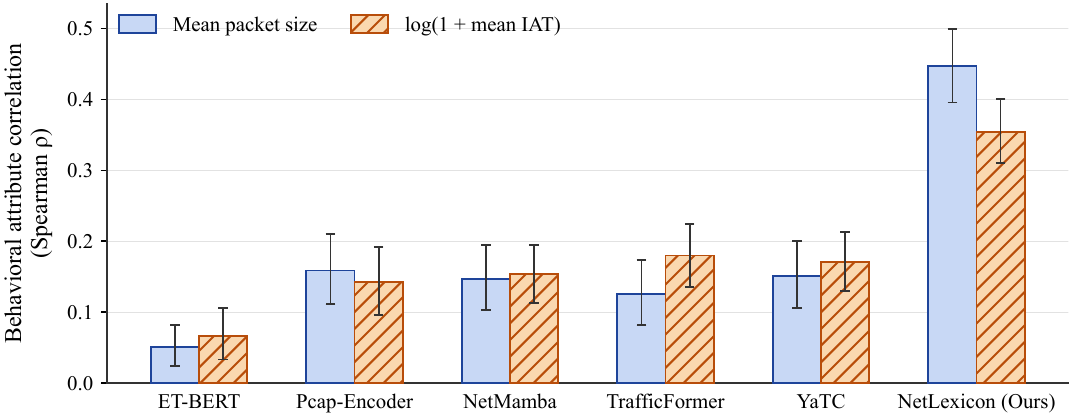}
  \caption{Association between representation distances and differences in packet size and inter-arrival time on CipherSpectrum.}
  \label{fig:tsne-compare}
  \vspace{-1.5em}
\end{figure}

Using the same 16,441 CipherSpectrum test flows, we compute pairwise embedding distances within each application class and correlate them with differences in the corresponding traffic attribute. We then average the Spearman correlations across classes. This measures whether representations distinguish behavioral variation among flows with the same application label. Figure~\ref{fig:tsne-compare} shows that \sys achieves correlations of 0.4476 for packet size and 0.3537 for inter-arrival time, approximately \(2.82\times\) and \(1.97\times\) those of the strongest baselines. Error bars denote 95\% confidence intervals from 10,000 bootstrap resamples over classes. These results show that the learned representation retains size and timing information beyond the separation of application classes.

\begin{figure*}[t]
  \centering
  \includegraphics[width=\linewidth]{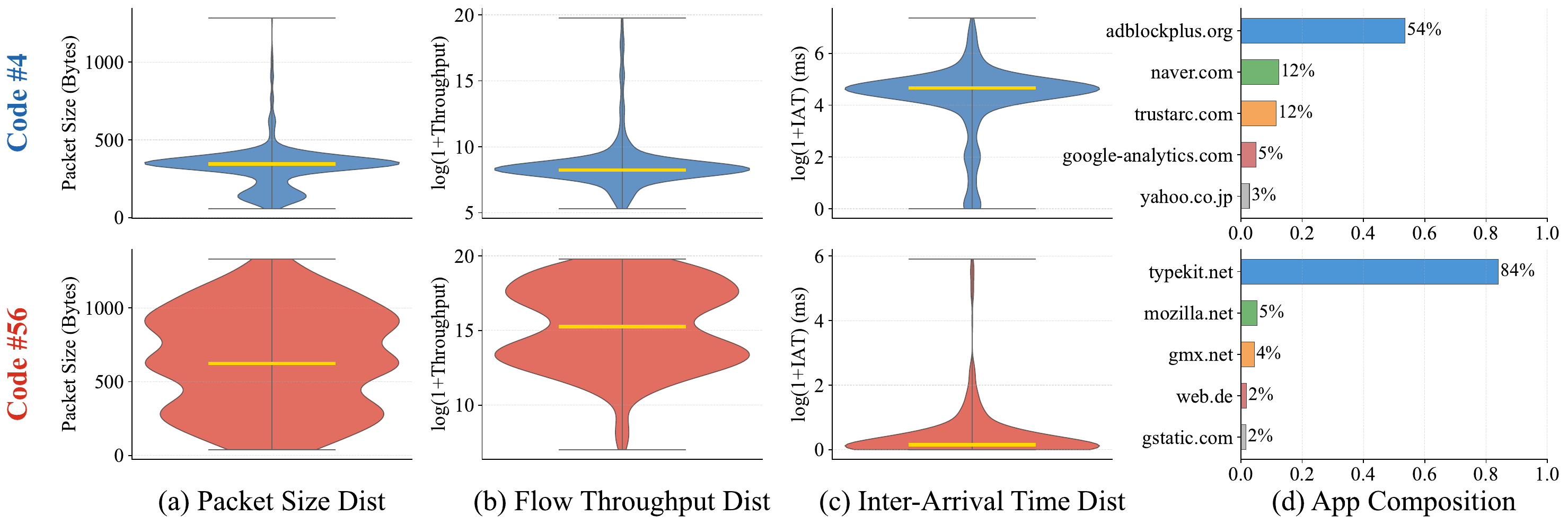}
  \caption{Traffic profiles of two representative codewords in the fine-tuned CipherSpectrum model~\cite{wickramasinghe2025sok}.}
  \label{fig:codebook_case}
  \vspace{-0.8em}
\end{figure*}

\noindent\textbf{Representative Codewords.}
Figure~\ref{fig:codebook_case} examines two codewords from the fine-tuned CipherSpectrum model. For each codeword, we summarize packet size, throughput, and inter-arrival time over its assigned windows and inspect the corresponding application composition. This connects the statistical profile of a state with the traffic sources in which it appears.

\begingroup
\emergencystretch=2em
Code~\#4 groups windows with medium packet sizes, relatively low throughput, and longer inter-arrival times, forming a profile of sparse exchanges. More than half of its assigned windows come from \texttt{adblockplus.org}, with additional windows from \texttt{naver.com}, \texttt{trustarc.com}, \texttt{google-analytics.com}, and \texttt{yahoo.co.jp}. Code~\#56 exhibits larger packet sizes, higher throughput, and shorter inter-arrival times, forming a profile of sustained transfer. Its dominant application is \texttt{typekit.net}, with additional contributions from \texttt{mozilla.net}, \texttt{gmx.net}, \texttt{web.de}, and \texttt{gstatic.com}.\par\endgroup

These examples illustrate two distinct communication patterns represented by the lexicon. Their behavioral interpretation follows from the measured size, throughput, and timing profiles, while the application composition shows that each state is reused across traffic from multiple sources. Together with the quantitative comparison, they provide evidence that the learned states organize windows around observable communication behavior.

\begin{figure*}[t]
  \centering
  \includegraphics[width=\linewidth]{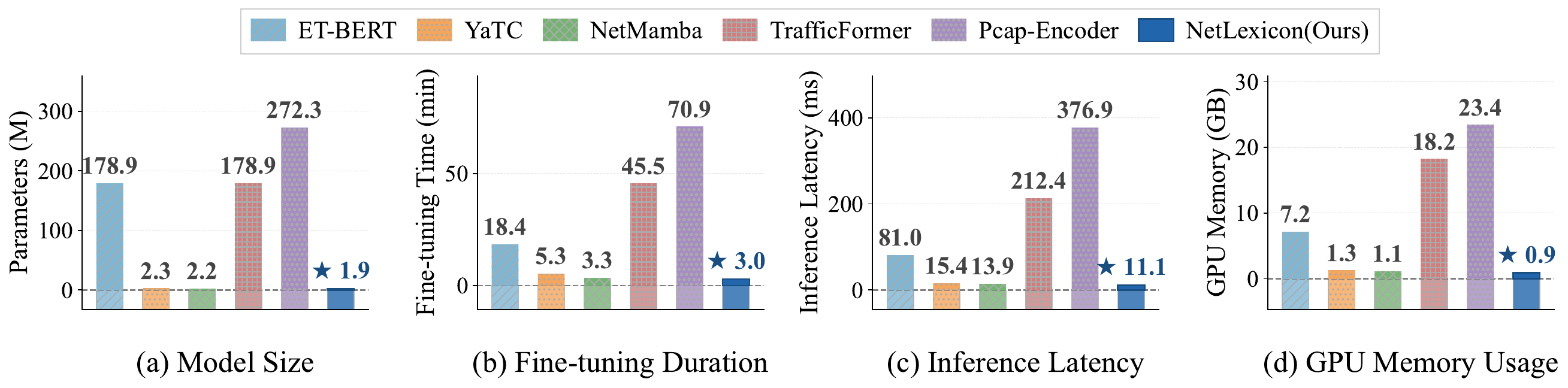}
  \caption{Comparison of model size, fine-tuning time per epoch, inference latency, and GPU memory usage.}
  \label{fig:efficiency}
  \vspace{-1em}
\end{figure*}

\vspace{-0.5em}

\subsection{Efficiency and Overhead}

Figure~\ref{fig:efficiency} compares parameter count, fine-tuning time, inference latency, and GPU memory usage. These measurements characterize both the cost of adapting a pretrained model and the resources required for inference.

\sys contains 1.9M parameters and requires 3.0 minutes per fine-tuning epoch, with 4,106 batches in each epoch. Its epoch time is up to \(23.6\times\) lower than the evaluated baselines, supporting efficient adaptation to downstream tasks. At an inference batch size of 32, the measured latency is 11.1 ms and GPU memory usage is 0.9 GB. Together with the classification results, these measurements show that \sys achieves strong performance with a compact model and modest computational requirements.

\section{Discussion}
\label{sec:discussion}

\noindent\textbf{Benefits and Tradeoffs of Discretization.}
Our results suggest that a compact vocabulary of traffic states can preserve information useful across several classification tasks. The value of this vocabulary lies in sharing behavioral patterns across flows. For example, sustained transfer and intermittent exchange can occur in multiple applications, allowing a shared state to capture common communication behavior while downstream supervision determines its relevance to a particular task. This perspective places traffic semantics at the level of interaction patterns and treats the lexicon as a reusable summary of these behaviors.

The granularity of this representation introduces a tradeoff. Sharing codewords can consolidate similar interactions and reduce sensitivity to minor variations, while also merging distinctions that matter for fine-grained classification. The codebook capacity and the context used to assign states therefore jointly determine which behavioral differences remain visible. STP and SFA guide this process through predictive and statistical signals. Their effectiveness should be assessed through both downstream performance and the behavioral structure of learned states, alongside careful evaluation of dataset artifacts and shortcut learning~\cite{wickramasinghe2025sok,zhao2025sweet,arp2022and,jacobs2022ai}.

\noindent\textbf{Limitations and Future Work.}
Reusing behavioral states across environments depends on how consistently the underlying interactions appear in observable traffic. Changes in network conditions or protocol implementations can alter packet sizes and timing, motivating adaptation of both the encoder and the codebook. Future work could explore how to incorporate new behaviors while preserving useful existing states. Larger contexts or hierarchical representations could further connect local window patterns into longer interaction sequences.

Deployment requires considering the full encoding pipeline. A compact codebook makes state storage and lookup inexpensive, while producing contextual window representations still requires neural computation. Prior work on programmable switches and network hardware offers useful directions for efficient traffic analysis~\cite{michel2021programmable,zhuo2023netbeacon,xie2022mousika,liu2021jaqen,yan2024brain,OSDIfpgaids,alcoz2025everything}. An important next step is to investigate whether the learned state assignments can supervise a smaller encoder or a hardware-compatible predictor, jointly addressing inference cost and behavioral fidelity~\cite{michel2021programmable,yan2024brain,paolucci2024cascaded,zhang2025pegasus,gao2025fenix}.

\section{Conclusion}
\label{sec:conclusion}

In this paper, we presented \sys, a traffic pretraining framework that learns traffic semantics through a compact lexicon of reusable behavioral states. Motivated by evolving applications and deployments under limited labeling budgets, \sys models how end-to-end communication behavior evolves over time. STP makes the learned states predictive of future traffic behavior, while SFA grounds them in window statistics. Across downstream traffic classification tasks, \sys improves Macro-F1 by up to 25.5 percentage points over strong baselines, reduces fine-tuning time per epoch by up to 23.6$\times$, and learns interpretable states aligned with meaningful communication behaviors. These results support discrete behavioral modeling as a promising approach to efficient and interpretable traffic analysis.
\bibliographystyle{ACM-Reference-Format}
\bibliography{main}
\appendix

\section{Feature Definitions}
\label{app:feature_tables}
This appendix summarizes the packet-level input features used by \sys and the window-level statistical summaries used by SFA. The two groups are presented separately because they play different roles in the model. Packet-level features serve as sequential inputs to the causal encoder and also act as regression targets in STP. In contrast, window-level statistics provide the auxiliary summary view used by SFA.

\noindent\textbf{Packet-level Features.}
Although the raw dataset contains a larger packet-level feature set, \sys uses only seven dimensions as packet-level model inputs. These features are selected to capture packet size, directionality, timing, and lightweight transport-level progression signals, while avoiding reliance on a large number of protocol-specific or deployment-specific fields.

\begin{table}[!ht]
\centering
\caption{Packet-level raw features used by \sys.}
\label{tab:packet_features_appendix}
\small
\setlength{\tabcolsep}{3.5pt}
\renewcommand{\arraystretch}{1.1}
\begin{tabularx}{\linewidth}{c l X c}
\toprule
Idx & Feature & Definition & Unit \\
\midrule
0 & Packet Size & Packet length, clipped to 1500. & B \\
1 & Direction & 0 for the first-packet direction, and 1 for the reverse direction. & -- \\
2 & Inter-Arrival Time & Time since the previous packet; the first packet is 0. & ms \\
3 & TCP Flags & Encoded TCP flags; set to -1 when TCP is unavailable. & -- \\
4 & $\Delta$IPID & Difference from the previous same-direction IP identifier; -1 if unavailable. & -- \\
5 & Relative Seq & TCP sequence number relative to the same-direction initial sequence number; -1 for non-TCP packets. & -- \\
6 & Relative Ack & TCP acknowledgment number relative to the opposite-direction initial sequence number; -1 if unavailable. & -- \\
\bottomrule
\end{tabularx}
\vspace{-1.5em}
\end{table}

In implementation, the encoder does not consume the raw 7-dimensional packet feature vector directly. Instead, each packet token is represented by a 14-dimensional input vector, formed by concatenating a normalized 7-dimensional feature vector with a 7-dimensional missing-value indicator. Let $\mathbf{x}_t^{\mathrm{raw}} \in \mathbb{R}^{7}$ denote the raw packet feature vector of packet token $t$. The final packet input is
\begin{equation}
\mathbf{x}_t = [\,\tilde{\mathbf{x}}_t \,;\, \mathbf{m}_t\,] \in \mathbb{R}^{14},
\end{equation}
where $\tilde{\mathbf{x}}_t \in \mathbb{R}^{7}$ denotes the normalized feature vector and $\mathbf{m}_t \in \mathbb{R}^{7}$ denotes the corresponding missing-value mask. Raw missing or inapplicable values are encoded as $-1$. During preprocessing, such entries are replaced by zero in the normalized branch and are explicitly marked through the missing-value mask.

The packet-level normalization is feature-specific. Packet size is scaled by 1500, direction is kept in its original binary form, and TCP flags are scaled by 255. For heavy-tailed timing and transport-progress features, \sys applies logarithmic compression before standardization. Specifically, inter-arrival time is transformed by $\log(1+x)$ and then z-score normalized; relative sequence number and relative acknowledgment number are transformed by $\log(1+\max(0,x))$ and then z-score normalized. By contrast, $\Delta$IPID is directly z-score normalized. Formally,
\begin{align}
\tilde{x}^{(\mathrm{size})} &= x^{(\mathrm{size})} / 1500, \\
\tilde{x}^{(\mathrm{dir})} &= x^{(\mathrm{dir})}, \\
\tilde{x}^{(\mathrm{iat})} &=
\frac{\log(1 + x^{(\mathrm{iat})}) - \mu_{\mathrm{iat}}}
{\sigma_{\mathrm{iat}} + \epsilon}, \\
\tilde{x}^{(\mathrm{flag})} &= x^{(\mathrm{flag})} / 255, \\
\tilde{x}^{(\Delta \mathrm{IPID})} &=
\frac{x^{(\Delta \mathrm{IPID})} - \mu_{\mathrm{ipid}}}
{\sigma_{\mathrm{ipid}} + \epsilon}, \\
\tilde{x}^{(\mathrm{seq})} &=
\frac{\log(1 + \max(0, x^{(\mathrm{seq})})) - \mu_{\mathrm{seq}}}
{\sigma_{\mathrm{seq}} + \epsilon}, \\
\tilde{x}^{(\mathrm{ack})} &=
\frac{\log(1 + \max(0, x^{(\mathrm{ack})})) - \mu_{\mathrm{ack}}}
{\sigma_{\mathrm{ack}} + \epsilon},
\end{align}
where $\mu$ and $\sigma$ denote the corresponding normalization statistics, and $\epsilon$ is a small constant for numerical stability. If a feature is missing in the raw input, its normalized value is set to 0 and the corresponding mask entry is set to 1. For non-packet tokens, including \texttt{[SEP]}, \texttt{[WIN]}, and \texttt{PAD}, the feature input is set to the all-zero vector, while token type is represented separately.

\noindent\textbf{Window-level Statistical Summaries.}
For each valid window, \sys computes a 37-dimensional statistical summary over a sliding window of five consecutive bursts. These features are not used as sequential input tokens. Instead, they provide a window-level statistical view aligned with the quantized discrete state used by SFA. For clarity, we organize them by the type of traffic behavior they describe.

\noindent\textbf{Basic traffic statistics.}
This group captures coarse traffic volume and directional asymmetry within a window. It summarizes how much traffic appears in each direction, how many packets are observed, and how packet sizes behave at a basic aggregate level.

\begin{table}[!ht]
\centering
\caption{Window-level basic traffic statistics.}
\label{tab:window_basic_appendix}
\small
\setlength{\tabcolsep}{3.5pt}
\renewcommand{\arraystretch}{1.08}
\begin{tabularx}{\linewidth}{c l X c}
\toprule
Idx & Feature & Definition & Unit \\
\midrule
1  & Flow Duration & Sum of inter-arrival times in the window. & ms \\
2  & Total Fwd Packets & Number of forward packets. & count \\
3  & Total Bwd Packets & Number of backward packets. & count \\
4  & Total Fwd Bytes & Total forward packet bytes. & B \\
5  & Total Bwd Bytes & Total backward packet bytes. & B \\
6  & Fwd Packet Length Mean & Mean forward packet size. & B \\
7  & Bwd Packet Length Mean & Mean backward packet size. & B \\
8  & Fwd Packet Length Std & Std. of forward packet sizes. & B \\
9  & Bwd Packet Length Std & Std. of backward packet sizes. & B \\
10 & Flow Bytes/s & Total bytes divided by window duration. & B/s \\
\bottomrule
\end{tabularx}
\vspace{-1.5em}
\end{table}

\noindent\textbf{Packet-length statistics.}
This group describes the shape of the packet-size distribution within the window. Beyond first- and sec\-ond-order statistics, it also includes extrema and skewness, allowing the summary to reflect whether traffic is concentrated around a narrow size range or exhibits more asymmetric patterns.

\begin{table}[!ht]
\centering
\caption{Window-level packet-length statistics.}
\label{tab:window_length_appendix}
\small
\setlength{\tabcolsep}{3.5pt}
\renewcommand{\arraystretch}{1.08}
\begin{tabularx}{\linewidth}{c l X c}
\toprule
Idx & Feature & Definition & Unit \\
\midrule
11 & Packet Length Mean & Mean packet size over all packets. & B \\
12 & Packet Length Std & Std. of packet sizes over all packets. & B \\
13 & Packet Length Variance & Variance of packet sizes. & B$^2$ \\
14 & Packet Length Max & Maximum packet size. & B \\
15 & Packet Length Min & Minimum packet size. & B \\
16 & Fwd Packet Length Max & Maximum forward packet size. & B \\
17 & Bwd Packet Length Max & Maximum backward packet size. & B \\
18 & Packet Length Skewness & Skewness of the packet-size distribution. & -- \\
\bottomrule
\end{tabularx}
\vspace{-1.5em}
\end{table}

\noindent\textbf{Timing statistics.}
This group characterizes the temporal structure of the interaction. In addition to flow-level inter-arrival-time summaries, it includes direction-specific timing statistics so that the forward and backward directions can exhibit distinct pacing behaviors within the same window.

\begin{table}[!ht]
\centering
\caption{Window-level timing statistics.}
\label{tab:window_timing_appendix}
\small
\setlength{\tabcolsep}{3.5pt}
\renewcommand{\arraystretch}{1.08}
\begin{tabularx}{\linewidth}{c l X c}
\toprule
Idx & Feature & Definition & Unit \\
\midrule
19 & Flow IAT Mean & Mean inter-arrival time over all packets. & ms \\
20 & Flow IAT Std & Std. of inter-arrival times over all packets. & ms \\
21 & Flow IAT Max & Maximum inter-arrival time. & ms \\
22 & Flow IAT Min & Minimum inter-arrival time. & ms \\
23 & Fwd IAT Mean & Mean IAT over adjacent forward packets. & ms \\
24 & Fwd IAT Std & Std. of IAT over adjacent forward packets. & ms \\
25 & Bwd IAT Mean & Mean IAT over adjacent backward packets. & ms \\
26 & Bwd IAT Std & Std. of IAT over adjacent backward packets. & ms \\
\bottomrule
\end{tabularx}
\vspace{-1.5em}
\end{table}

\noindent\textbf{TCP control statistics.}
This group summarizes transport-level events within the window, including setup, acknowledgment, reset, and teardown behavior. It provides a compact representation of control-plane activity without requiring explicit protocol parsing.

\begin{table}[!ht]
\centering
\caption{Window-level TCP control statistics.}
\label{tab:window_tcp_appendix}
\small
\setlength{\tabcolsep}{3.5pt}
\renewcommand{\arraystretch}{1.08}
\begin{tabularx}{\linewidth}{c l X c}
\toprule
Idx & Feature & Definition & Unit \\
\midrule
27 & FIN Flag Count & Number of packets with FIN set. & count \\
28 & SYN Flag Count & Number of packets with SYN set. & count \\
29 & RST Flag Count & Number of packets with RST set. & count \\
30 & PSH Flag Count & Number of packets with PSH set. & count \\
31 & ACK Flag Count & Number of packets with ACK set. & count \\
32 & URG Flag Count & Number of packets with URG set. & count \\
\bottomrule
\end{tabularx}
\vspace{-1.5em}
\end{table}

\noindent\textbf{Flow-behavior statistics.}
This final group provides a coarse summary of burstiness and activity patterns. It complements the previous groups by describing how traffic intensity varies over time and how active and idle periods are distributed within the window.

\begin{table}[!ht]
\centering
\caption{Window-level flow-behavior statistics.}
\label{tab:window_behavior_appendix}
\small
\setlength{\tabcolsep}{3.5pt}
\renewcommand{\arraystretch}{1.08}
\begin{tabularx}{\linewidth}{c l X c}
\toprule
Idx & Feature & Definition & Unit \\
\midrule
33 & Fwd Packets/s & Forward packet rate. & packets/s \\
34 & Bwd Packets/s & Backward packet rate. & packets/s \\
35 & Active Mean & Mean duration of active periods. & s \\
36 & Active Std & Std. of active-period duration. & s \\
37 & Idle Mean & Mean duration of idle periods. & s \\
\bottomrule
\end{tabularx}
\vspace{-1.5em}
\end{table}

Taken together, these groups form the window-level statistical view used by SFA. They help anchor the learned discrete states to stable and observable traffic behaviors.

\section{Implementation Details}
\label{app:implementation}

We initialize the baselines from their publicly released pretrained checkpoints and retain their respective input formats. ET-BERT~\cite{et-bert} and TrafficFormer~\cite{zhou2025trafficformer} represent each flow using five packets with 64 bytes per packet. YaTC~\cite{zhao2023yet} and NetMamba~\cite{wang2024netmamba} convert each flow into a \(40\times40\) image following their preprocessing pipelines. Pcap-Encoder~\cite{zhao2025sweet} uses individual packets as samples in our reproduction. For \sys, each packet is represented by seven normalized features and seven missing indicators, with separate embeddings for token types and positions. Padding positions are masked from attention and pretraining losses. The STP prediction heads and SFA projection heads are used during pretraining. Downstream classification pools the contextual \texttt{[WIN]} representations and applies a task-specific classification head. We first train this head with the backbone frozen, then fine-tune the model jointly. An AdamW optimizer is instantiated over the trainable parameters at each stage, with the head-only stage using ten times the base learning rate. All windows from a flow inherit its dataset split.

\section{Additional Codebook Entries}
\label{app:more_codebook_cases}
Table~\ref{tab:top50_codebook_entries} reports the 50 most frequently activated codebook entries on the CipherSpectrum~\cite{wickramasinghe2025sok} test split after fine-tuning. For each code, we show its assignment count, key traffic statistics, dominant application, and a heuristic behavioral tag. These entries exhibit recurring patterns such as TCP startup, short download-heavy bursts, balanced exchanges, and longer transfers with idle gaps. Their assignment counts indicate how frequently each state is reused, while the statistical profiles describe the communication behavior associated with it. Together, these observations provide a concrete view of how the learned codebook organizes traffic windows into interpretable behavioral patterns across the evaluated traces.

\begin{table*}[t]
\centering
\caption{Top 50 active codebook entries from the fine-tuned CipherSpectrum model on the test split.}
\label{tab:top50_codebook_entries}
\small
\begin{tabular*}{\textwidth}{@{\extracolsep{\fill}}l c c c c c c c c c p{4.3cm}@{}}
\toprule
Code & Count & Dur. & Fwd & Bwd & FwdB & BwdB & SYN & IAT & Dominant APP$^{1}$ & Tag$^{2}$ \\
\midrule
91  & 10049 & 64   & 3 & 4 & 191  & 4411 & 0.0 & 11  & ctfassets        & short + DL-heavy \\
45  & 7561  & 44   & 3 & 4 & 255  & 4401 & 0.0 & 7   & googletagmanager & short + DL-heavy \\
34  & 4032  & 207  & 3 & 4 & 160  & 4488 & 0.0 & 39  & gstatic          & medium + DL-heavy \\
33  & 4000  & 25   & 5 & 4 & 1774 & 1623 & 2.0 & 3   & hubspot          & startup + balanced \\
6   & 2447  & 113  & 5 & 4 & 1731 & 1640 & 2.0 & 15  & google-analytics & startup + balanced \\
66  & 2341  & 27   & 4 & 3 & 835  & 1582 & 2.0 & 5   & ctfassets        & startup + DL-heavy \\
126 & 2222  & 122  & 3 & 4 & 348  & 3846 & 0.0 & 21  & arin             & short + DL-heavy \\
121 & 1993  & 10   & 4 & 3 & 1155 & 1639 & 1.9 & 2   & twitter          & startup + short \\
20  & 1897  & 290  & 3 & 4 & 447  & 5015 & 0.0 & 45  & adblockplus      & medium + DL-heavy \\
32  & 1743  & 612  & 4 & 4 & 476  & 2833 & 0.0 & 89  & trustarc         & medium + DL-heavy \\
117 & 1671  & 53   & 4 & 4 & 699  & 2213 & 0.9 & 9   & garmin           & short + startup \\
36  & 1607  & 424  & 4 & 3 & 1187 & 1604 & 2.0 & 65  & arin             & startup + balanced \\
103 & 1553  & 321  & 3 & 4 & 359  & 3435 & 0.0 & 53  & segment          & medium + DL-heavy \\
120 & 1528  & 290  & 3 & 4 & 224  & 3787 & 0.0 & 46  & googletagmanager & medium + DL-heavy \\
127 & 1518  & 1073 & 3 & 4 & 391  & 3821 & 0.0 & 197 & arin             & idle-gap + DL-heavy \\
54  & 1450  & 30   & 3 & 3 & 194  & 3187 & 0.0 & 5   & ctfassets        & short + DL-heavy \\
59  & 1405  & 26   & 3 & 5 & 461  & 1326 & 0.0 & 4   & mozilla          & short + response-heavy \\
77  & 1330  & 478  & 4 & 4 & 1075 & 2745 & 0.0 & 66  & arin             & medium + balanced \\
69  & 1316  & 169  & 3 & 4 & 281  & 3895 & 0.1 & 25  & flipboard        & short + DL-heavy \\
48  & 1271  & 34   & 4 & 4 & 322  & 2369 & 0.0 & 5   & typekit          & short + DL-heavy \\
101 & 1266  & 910  & 3 & 4 & 359  & 2242 & 0.0 & 147 & robinhood        & idle-gap + DL-heavy \\
107 & 1244  & 65   & 3 & 4 & 251  & 3070 & 0.0 & 10  & steamstatic      & short + DL-heavy \\
21  & 1232  & 243  & 3 & 4 & 408  & 3792 & 0.0 & 34  & garmin           & medium + DL-heavy \\
68  & 1220  & 84   & 3 & 4 & 476  & 4041 & 0.0 & 12  & steamstatic      & short + DL-heavy \\
49  & 1149  & 7    & 3 & 5 & 170  & 6559 & 0.0 & 1   & gstatic          & dense + DL-heavy \\
104 & 1127  & 781  & 4 & 4 & 735  & 2939 & 0.0 & 117 & getpocket        & idle-gap + DL-heavy \\
100 & 1120  & 115  & 4 & 4 & 264  & 2684 & 0.0 & 16  & yimg             & short + DL-heavy \\
90  & 1067  & 98   & 3 & 5 & 367  & 5899 & 0.0 & 14  & getpocket        & short + DL-heavy \\
19  & 1013  & 17   & 4 & 4 & 625  & 614  & 0.0 & 2   & hs-scripts       & dense + balanced \\
113 & 929   & 482  & 3 & 4 & 437  & 3993 & 0.0 & 71  & cookielaw        & medium + DL-heavy \\
43  & 924   & 118  & 3 & 4 & 257  & 3893 & 0.0 & 19  & googletagmanager & short + DL-heavy \\
102 & 876   & 592  & 3 & 4 & 369  & 3068 & 0.0 & 94  & googletagmanager & medium + DL-heavy \\
92  & 779   & 33   & 3 & 4 & 194  & 3346 & 0.0 & 5   & xnxx-cdn         & short + DL-heavy \\
70  & 773   & 48   & 3 & 4 & 183  & 4696 & 0.0 & 9   & naver            & short + DL-heavy \\
84  & 754   & 782  & 3 & 4 & 279  & 3395 & 0.0 & 125 & cookielaw        & idle-gap + DL-heavy \\
22  & 701   & 110  & 5 & 4 & 727  & 1233 & 0.0 & 13  & googletagmanager & short + balanced \\
40  & 695   & 555  & 4 & 3 & 1029 & 1684 & 1.8 & 85  & ubuntu           & startup + balanced \\
0   & 686   & 247  & 4 & 4 & 887  & 2671 & 0.0 & 28  & wistia           & medium + balanced \\
108 & 660   & 250  & 3 & 4 & 300  & 3283 & 0.0 & 40  & coinbase         & medium + DL-heavy \\
28  & 659   & 979  & 4 & 4 & 1567 & 1960 & 1.1 & 128 & gmx              & startup + idle-gap \\
115 & 657   & 328  & 3 & 5 & 165  & 6621 & 0.0 & 55  & googletagmanager & medium + DL-heavy \\
46  & 648   & 532  & 3 & 4 & 202  & 4286 & 0.0 & 86  & segment          & medium + DL-heavy \\
82  & 646   & 720  & 3 & 3 & 551  & 3680 & 0.0 & 121 & getpocket        & idle-gap + DL-heavy \\
112 & 605   & 166  & 4 & 3 & 773  & 1589 & 1.7 & 27  & like-video       & startup + balanced \\
111 & 588   & 50   & 5 & 4 & 1912 & 1712 & 1.9 & 8   & typekit          & startup + balanced \\
10  & 559   & 13   & 2 & 4 & 124  & 5851 & 0.0 & 2   & googletagmanager & dense + DL-heavy \\
\underline{4}  & 545 & 774 & 4 & 3 & 919 & 1509 & 1.3 & 120 & \underline{adblockplus} & \underline{startup + idle-gap} \\
\underline{56} & 532 & 33  & 3 & 5 & 275 & 4982 & 0.0 & 5   & \underline{typekit}     & \underline{short + DL-heavy} \\
61  & 524   & 192  & 3 & 4 & 267  & 4933 & 0.0 & 33  & coinbase         & medium + DL-heavy \\
83  & 514   & 108  & 4 & 4 & 1505 & 1041 & 0.0 & 14  & wistia           & short + balanced \\
\bottomrule
\end{tabular*}
\parbox{\textwidth}{\raggedright\footnotesize
\textit{Note:} 
The underlined rows correspond to the representative codewords analyzed in Figure~\ref{fig:codebook_case}.\\
$^{1}$\textit{Dominant APP} denotes the most frequent application among windows assigned to each code and is descriptive only rather than a semantic label; it does not imply that the code is exclusive to that application, as each code may still be shared across multiple applications.\\
$^{2}$\textit{Tag} is a coarse rule-based summary derived from aggregate statistics: 
\textit{startup} indicates elevated SYN activity, 
\textit{DL-heavy} indicates substantially larger backward than forward bytes, 
\textit{balanced} indicates comparable bidirectional byte volume, 
\textit{idle-gap} indicates relatively large duration or mean inter-arrival time, 
\textit{dense} indicates very small mean inter-arrival time, 
and \textit{response-heavy} indicates a larger backward packet count with modest byte volume. \\
}
\end{table*}

\end{document}